\documentclass[final,number,sort&compress,5p,times,twocolumn]{elsarticle} 

\usepackage[T1]{fontenc}
\usepackage{graphicx}
\usepackage{amsmath}
\usepackage[colorlinks]{hyperref}
\usepackage[nameinlink,noabbrev]{cleveref}

\journal{Nuclear Instruments and Methods in Physics Research Section A}

\usepackage{etoolbox}
\makeatletter
    \patchcmd{\tnotemark}{\ding{73}}{*}{}{\@latex@error{Failed to path \string\tnotemark\space for \string\ding{73}}}
\patchcmd{\tnotemark}{\ding{73}\ding{73}}{\dag}{}{\@latex@error{Failed to path \string\tnotemark\space for \string\ding{73}\string\ding{73}}}
    \patchcmd{\tnotetext}{\ding{73}}{*}{}{\@latex@error{Failed to path \string\tnotetext\space for \string\ding{73}}}
\patchcmd{\tnotetext}{\ding{73}\ding{73}}{\dag}{}{\@latex@error{Failed to path \string\tnotetext\space for \string\ding{73}\string\ding{73}}}
\makeatother 

\usepackage{etoolbox}
\makeatletter
\patchcmd{\ps@pprintTitle}
  {Preprint submitted to}
  {Accepted manuscript by}
  {}{}
\makeatother

\begin{document}


\begin{frontmatter}

\author[add1]{S.~Sailer\corref{cor}}
\ead{simon.sailer@mpi-hd.mpg.de}
\author[add1]{F.~Werner}
\author[add1]{G.~Hermann}
\author[add1]{M.~Barcelo}
\author[add1]{C.~Bauer}
\author[add3]{S.~Bernhard}
\author[add3]{M.~Biegger}
\author[add5]{F.~Canelli}
\author[add3]{M.~Capasso}
\author[add3]{S.~Diebold}
\author[add3]{F.~Eisenkolb}
\author[add4]{S.~Eschbach}
\author[add5]{D.~Florin}
\author[add1]{C.~Föhr}
\author[add4]{S.~Funk}
\author[add5]{A.~Gadola}
\author[add1]{F.~Garrecht}
\author[add4]{I.~Jung}
\author[add4]{O.~Kalekin}
\author[add2]{C.~Kalkuhl}
\author[add1]{T.~Kihm}
\author[add4]{R.~Lahmann}
\author[add4]{M.~Pfeifer}
\author[add4]{G.~Principe}
\author[add2]{G.~Pühlhofer}
\author[add1]{S.~Pürckhauer}
\author[add3]{O.~Reimer}
\author[add2]{A.~Santangelo}
\author[add2]{M.~Scalici}
\author[add2]{T.~Schanz}
\author[add1]{T.~Schwab}
\author[add5]{S.~Steiner}
\author[add5]{U.~Straumann}
\author[add2]{C.~Tenzer}
\author[add5]{A.~Vollhardt}
\author[add5]{D.~Wolf}

\cortext[cor]{Corresponding author}

\address[add1]{Max-Planck Institut für Kernphysik, Heidelberg (Germany)}
\address[add2]{Eberhard Karls Universität Tübingen (Germany)}
\address[add3]{Leopold-Franzens Universität Innsbruck (Austria)}
\address[add4]{Friedrich-Alexander-Universität Erlangen-Nürnberg (Germany)}
\address[add5]{Universität Zürich (Switzerland) }


\begin{abstract}
FlashCam is a camera proposed for the medium-sized telescopes of the Cherenkov Telescope Array (CTA).
We compare camera trigger rates obtained from measurements with the camera prototype in the laboratory and Monte-Carlo simulations, when scanning the parameter space of the fully-digital trigger logic and the intensity of a continuous light source mimicking the night sky background (NSB) during on-site operation.
The comparisons of the measured data results to the Monte-Carlo simulations are used to verify the FlashCam trigger logic and the expected trigger performance.

\end{abstract}


\end{frontmatter}

\section{Introduction}
FlashCam \cite{Puehl:2016} implements a fully-digital trigger processing and readout based on FADCs and FPGAs.
The signals of the 1758~photomultiplier~tubes (PMTs) are sampled continuously at a rate of 250~MS/s, and upon a trigger decision the digitised waveforms are sent via Ethernet to the data-acquisition server.
The trigger firmware logic has also been implemented in a software framework, which, when applied to the read-out traces, computes the same trigger signals as the camera electronics.
This allows a \textit{bit-exact} confirmation of the digital processing chain and an emulation of the trigger over the full range of trigger parameters using measured or simulated input data.

The influence of Poissonian fluctuations of the night sky background illumination and the properties of the PMTs on the camera trigger rates have been investigated using Monte-Carlo simulations.
We also show comparisons of these simulations with camera trigger rates derived from measured data, taken with the fully equipped prototype in the laboratory, by simulating continuous background light using an LED.

\subsection{Fully-digital topological trigger forming}
The FlashCam trigger logic is realised as a digital sum trigger, which is sensitive to local coincidences in time $\mathcal{O}$(ns) and space.
The camera electronics computes the full trigger chain continuously on the digitised samples.
The first stage is a channel-wise differentiating filter.
These differentiated signals are then scaled down because of bandwidth limitations and clipped to a maximum value.
The clipping also prevents triggering on large single-pixel signals such as PMT-afterpulses.

In the last stage, a sample-wise summation of the clipped signals over 9 neighbouring pixels is carried out each time-step for each of the 588 predefined, overlapping pixel patches, ensuring a homogeneous and seamless coverage over the camera plane.
A camera readout is triggered if one of these patch sums exceeds a predefined threshold.

\section{Monte-Carlo simulations vs Measurements}
Monte-Carlo simulations of the camera have been produced using the sim\_telarray simulation package~\cite{Bernloehr:2008} resulting in full-trace simulations of 1 s of digitisation of all camera pixels per simulation configuration.
The simulations were produced for two camera configurations (7- and 8-dynode PMTs), differing in their afterpulsing probability distributions and their pulse shapes, and with two NSB intensities each (300~MHz and 1.2~GHz). These intensities represent the lower and upper end of the NSB conditions on site.

The camera prototype, equipped with an even mixture of 7- and 8-dynode PMTs 12-pixel modules, was used to take randomly triggered data, while being illuminated with an LED, reading out events with a trace length of 3900 samples (15.6 $\mu s$), until also 1 s of digitised data was accumulated.
Both types of dataset were fed into the FlashCam trigger emulation framework to calculate the final camera trigger rates. The camera trigger rates computed by this framework have been compared to camera trigger rate measurements with the camera prototype and shown to match within the statistical uncertainties.

\begin{figure}[t]
\centering
\includegraphics[width=0.99\linewidth]{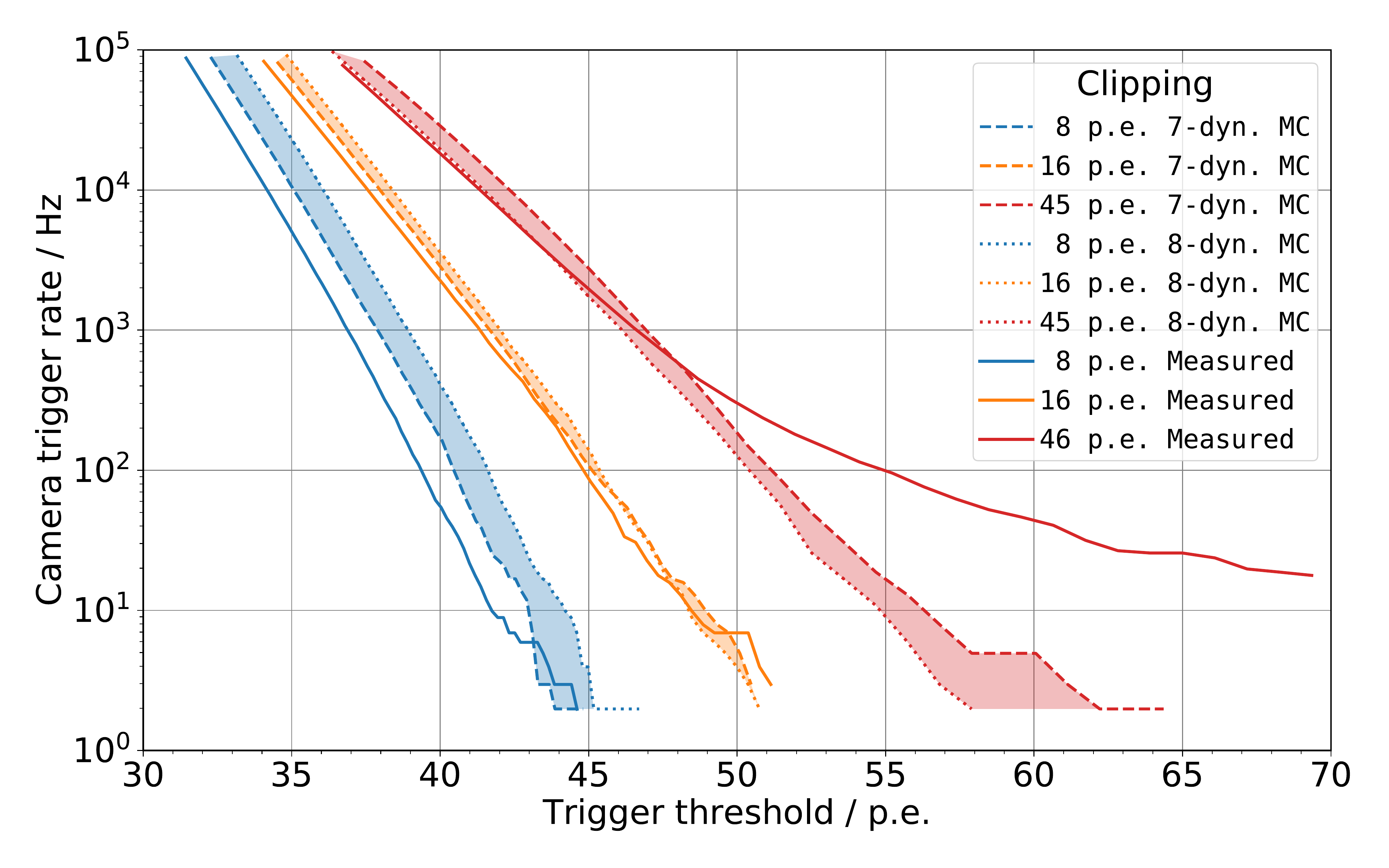}
\caption{Comparison of camera trigger rates from measured camera input data (with both 7- \textit{and} 8- dynode PMTs) and Monte-Carlo simulations (7- \textit{or} 8-dynode PMTs).
The shaded regions are bordered by the results of the simulation, while the solid lines show the results from measured input data.
The simulated NSB is 300~MHz of p.e./pixel.}
\label{fig:Rates300}
\end{figure}

The comparison of camera trigger rates produced with measured and simulated input datasets at 300~MHz NSB is shown in \autoref{fig:Rates300}.
The coloured regions show the Monte-Carlo simulations for the two camera configurations, while the solid lines show the camera trigger rates obtained from measurements with the prototype.
The camera trigger rates for three separate settings of pixel clipping levels are coloured in red, orange and blue.
The region of interest is at the planned nominal operation rate $\mathcal{O}($10 kHz), with a very good match for clipping levels of 16 or 45 p.e, while the largest difference in threshold is for a clipping level of 8~p.e.
This is due to the slightly conservative single photoelectron resolution used in the simulations.
Nevertheless, this difference is less than 5\% in trigger threshold for trigger rates $>$500 Hz for all clipping levels.
The relative trigger rate between 7- and 8-dynode simulations changes between low (8~p.e.) and high (45~p.e.) clipping levels, which can be explained by a higher afterpulsing probability ($1.2 \cdot 10^{-4}$ for pulses >~4~p.e.)\ of the 7-dynode PMTs compared to the 8-dynode PMTs ($1.0 \cdot 10^{-4}$ for pulses >~4~p.e.). In contrast the rise time of the 7-dynode pulses is shorter than the 8-dynode rise time (5.6~ns versus~6.1~ns), reducing the coincidence time of NSB photoelectrons.

\begin{figure}[t]
\centering
\includegraphics[width=0.99\linewidth]{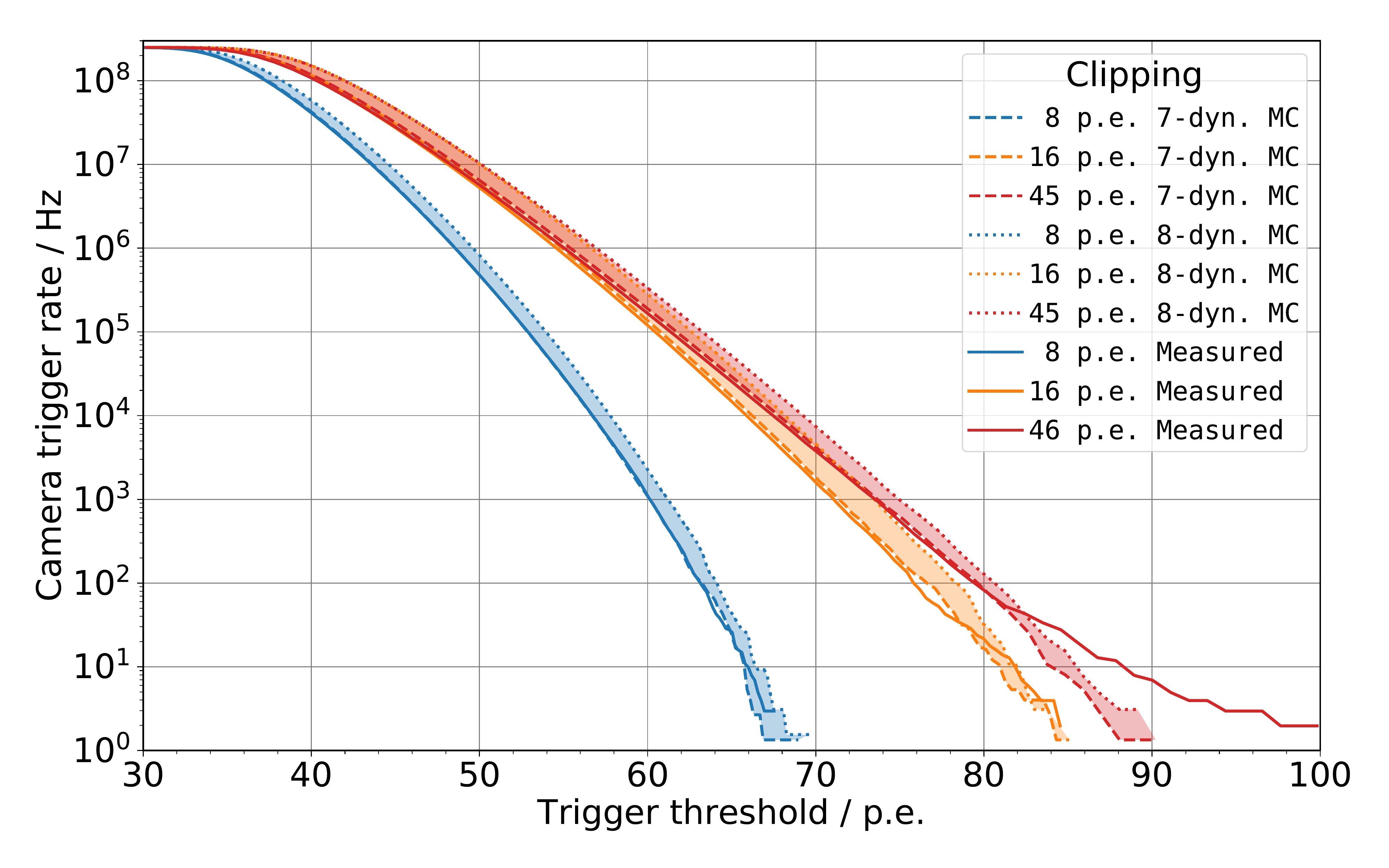}
\caption{Comparison of camera  trigger rates derived from measured camera input data (with both 7- \textit{and} 8- dynode PMTs) and Monte-Carlo simulations (7- \textit{or} 8-dynode PMTs).
The simulated NSB is 1.2~GHz of p.e./pixel}
\label{fig:Rates1200}
\end{figure}

The comparisons at 1.2~GHz NSB are presented in \autoref{fig:Rates1200}.
The simulations show a very good agreement with the measurements over the full range of trigger thresholds.
There is a small overall shift towards higher thresholds of the simulations compared to the rates from measured input data due to the single p.e. resolution used in the simulations.
The better match at low clippings (8~p.e.) compared to the 300~MHz case is due to photoelectron pile-up, as the expected number of photo electrons is higher with an NSB of 1.2~GHz (4.8~p.e.~/~pixel~/~sample).

The differences in camera trigger rates from all measurements (more pronounced in the 46~p.e.\ clipping levels) stem from atmospheric muons interacting with the plexiglass window or individual PMTs producing large signals.

\section{Conclusion}

The results presented here show a very good agreement of the Monte-Carlo simulations and the camera trigger rates obtained from measured data in the laboratory.
This also indicates that the Monte-Carlo input parameters describe the camera electronics and the PMT characteristics well.
Additionally, the two different simulated setups demonstrate the minor systematic differences at the 5\% level due to PMT parameter variation over the range of expected NSB levels.

\section*{Acknowledgements}

This work was conducted in the context of the CTA FlashCam Project.
We gratefully acknowledge financial support from the agencies and organizations listed here: \url{http://www.cta-observatory.org/consortium_acknowledgments}



\begin{thebibliography}{10}
\expandafter\ifx\csname url\endcsname\relax
  \def\url#1{\texttt{#1}}\fi
\expandafter\ifx\csname urlprefix\endcsname\relax\def\urlprefix{URL }\fi

\bibitem{Puehl:2016}
G.~P\"uhlhofer, et al., The CTA Consortium, The medium size telescopes of the
Cherenkov Telescope Array, in: Proc. of the 6th Intern. Symposium on High-Energy
Gamma-Ray Astronomy, 2016. arXiv:1610.02899.

\bibitem{Bernloehr:2008}
Konrad Bernl\"ohr, Simulation of imaging atmospheric Cherenkov telescopes with CORSIKA and sim\_telarray, Astroparticle Physics 30,
  (2008) 149--158.

\end{thebibliography}
\end{document}